\documentclass[
bibnotes,
amsmath,
amssymb,
preprint,
pre,
superscriptaddress
]{revtex4-2}
\usepackage{array}
\usepackage{subfigure}
\usepackage{graphicx}
\usepackage{dcolumn}
\usepackage{bm}
\usepackage{float}
\usepackage{ulem}
\usepackage{color}
\usepackage[colorlinks,
linkcolor = blue,
citecolor = blue,
urlcolor = blue,
bookmarks = false]{hyperref}
\usepackage[mathlines]{lineno}

\begin{document}
\renewcommand\arraystretch{2}

\title{Anomalous diffusion of self-align active particle in flow background}

\author{Ze-Hao Chen}
\affiliation{%
Institute of Computational Physics and Complex Systems, Lanzhou University, Lanzhou, Gansu 730000, China
}%
\author{Zhi-Xi Wu}%
\email{wuzhx@lzu.edu.cn}
\affiliation{%
Institute of Computational Physics and Complex Systems, Lanzhou University, Lanzhou, Gansu 730000, China
}%
\affiliation{%
Lanzhou Center for Theoretical Physics and Key Laboratory of Theoretical Physics of Gansu Province, Lanzhou University, Lanzhou, Gansu 730000, China
}%



\begin{abstract}Active particles (i.e., self-propelled particles or called microswimmers), different from passive Brownian particles, possess more complicated translational and angular dynamics, which can generate a series of anomalous transport phenomena. 
In this letter, we study the two-dimensional dynamics of a self-propelled pointlike particle with self-aligning property moving in Poiseuille flow. 
The results show the effective anomalous diffusion coefficient changes sharply with the change of temperature and speed of background Poiseuille flow. 
The relaxation property of moving speed and the position probability distribution function of particles is also obtained. 
The observation of several types of anomalous diffusion and normal diffusion regime indicates the self-aligning property may be universal and can be used as a reference for future experiments analysis and modeling.
\end{abstract}

\maketitle

\section{Introduction}
Self-propelled particles are ubiquitous models to describe the component of active systems in nature. 
According to the difference of angular dynamics, there are three basic self-propelled particles models, namely, active Brownian particles (ABPs), run-and-tumble particles (RTPs), and active Ornstein-Uhlenbeck particles (AOUPs)~\cite{RevModPhys.88.045006,D0SM00687D}. 
The mean squared displacement (MSD) of an isolated particle is the same for the three main models (i.e., the long time behavior of the MSD of the active particle is diffusively, ${\rm{MSD}}(t) \sim t$ and ballistically at short time, ${\rm{MSD}}(t) \sim t^2$) due to their pure noise angular dynamics, although the concrete form and mechanism of these noises may different.

However, active particles always moving in a complicated environment and possess more complex angular dynamics, resulting in anomalous diffusion or diffusion enhancement. 
For example, active particles moving in an environment full of obstacles or some external field~\cite{PhysRevE.65.031104,PhysRevLett.87.010602,PhysRevLett.93.120603,PhysRevLett.106.090602,PhysRevLett.123.128101,PhysRevResearch.2.022020,Spiechowicz2016,Spiechowicz2017,Spiechowicz_2019,PhysRevE.97.032603,Hanes_2012,UPPALURI20121162,C4CP03465A}, bacteria and pathogens~\cite{UPPALURI20121162,Secchi2020}, or artificial microswimmers (some could be used as drug deliverers) moving in the human body~\cite{choudhary_renganathan_pushpavanam_2019,PhysRevLett.105.268302,PhysRevLett.120.188101,IGNACIO2017486,GOMESFILHO201629}. 
Normal and anomalous diffusion is characterized by the power law evolution of ${\rm{MSD}}(t)$ \begin{equation}\label{eq:diffusion_def}
{\rm{MSD}}(t) \sim t^{\alpha}
\end{equation}
then normal diffusion is observed for $\alpha = 1$ and the anomalous diffusion can be classified as follows:
\begin{enumerate}
\item[(1)] subdiffusion for $0 < \alpha < 1$
\item[(2)] superdiffusion for $1 < \alpha < 2$
\item[(3)] ballistic diffusion for $\alpha = 2$
\item[(4)] superballistic or hyperdiffusion for $\alpha > 2$.
\end{enumerate}

Unraveling the underlying propulsion and angular dynamics mechanisms is essential to understand the behavior of microswimmers or to design biomimetics by transferring biological concepts to synthetic swimmers. 
One of the overarching challenges for this field is the distinction between the generic and specific properties and behavior of an active matter system. 
In other words, the question is what properties are universal and when specific mechanisms dominate their behavior.

As a basis of further research, this study is to examine the transportation property of an active particle with self-aligning property~\cite{PhysRevLett.110.208001,PhysRevLett.122.068002} moving in a semi-confine tube and thus interact with space varying circumstances.
The remainder of this paper is organized as follows. 
In the next section, we introduce our model as well as the corresponding quantifiers in detail. 
Numerical simulations and analyses are performed in the third section. 
Finally, a short conclusion is given. 

\section{\label{sec:model}The model}

\begin{figure}
\centering
\includegraphics[width=\columnwidth]{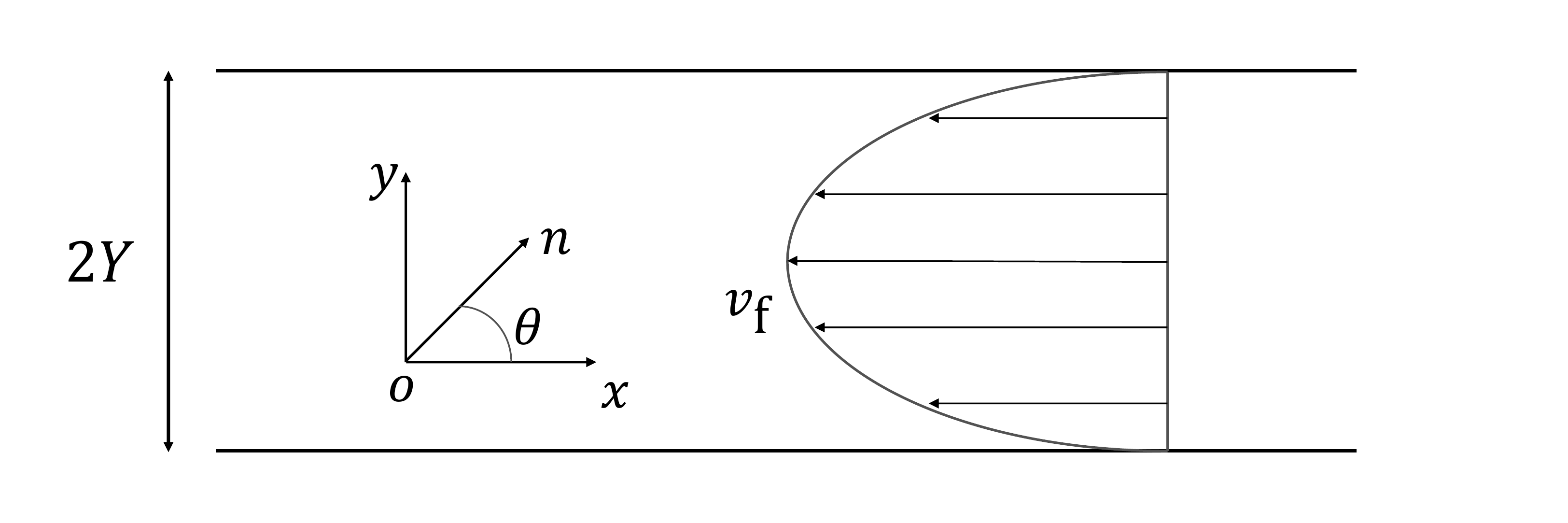}
\caption{An active particle moving in a 2D channel with channel width $Y$ where Poiseuille flow is imposed. The rectangular coordinate system with coordinate $(x, y)$ and the orientation angle $\theta$ to the $x$-axis.}
\label{fig:concept}
\end{figure}

We consider a microswimmer that is subject to self-propelled force $F$ and self-aligning torque with intensity $\zeta$ moves in a narrow 2D channel where Poiseuille flow is imposed, see fig.~\ref{fig:concept}. Here we omit the translational diffusion but consider the noise in the orientation dynamics, the evolution of velocity $\boldsymbol v$ and orientation $\boldsymbol n$ obey
\begin{equation}\label{eq:vrs_translation}
m\frac{\mathrm d\boldsymbol v}{\mathrm dt}=F\boldsymbol n-\gamma\left(\boldsymbol v-{\boldsymbol v}_{\mathrm f}\right),
\end{equation}
\begin{equation}\label{eq:vrs_rotation}
\tau\frac{\mathrm d\boldsymbol n}{\mathrm dt}=\zeta\left(\boldsymbol n\times\boldsymbol v\right)\times\boldsymbol n+\sqrt{2\alpha}\xi\left(t\right){\boldsymbol n}_\perp.
\end{equation}

Equation (\ref{eq:vrs_translation}) contains the mass of the microswimmer $m$, the friction coefficient $\gamma$, and the velocity of background Poiseuille flow ${\boldsymbol v}_{\mathbf f}$. 
The motivation for introducing such a second-order equation is that previous studies found that some special phenomenon in active systems occurs if and only if accounting for the inertia of active particles~\cite{PhysRevLett.123.228001,Spiechowicz_2019,SuneSimon2014}.
The overdamped orientation dynamic is adopted in eq.~(\ref{eq:vrs_rotation}), where the continuous cross product represents a self-aligning torque induced by an asymmetry dissipated force and we ignore the effect of the curl of the fluid on the particle orientation. 
This effect has been found in some systems such as vibrated polar disks~\cite{PhysRevLett.110.208001} and hexbug running in a vibrated parabolic dish~\cite{PhysRevLett.122.068002}. Besides, Gaussian noise $\xi\left(t\right)$ with zero mean and delta correlations $\left\langle\xi(t)\xi(t')\right\rangle=\delta(t-t')$ and the rotational diffusion coefficient $\alpha/\tau^2$ is contained. 
Finally, the velocity field of Poiseuille flow is given by a time-space separated variables form
\begin{equation}\label{eq:vrs_flow}
|{\boldsymbol v}_{\mathrm f}\left(y,\;t\right)|=\frac A{2\gamma}\left(Y^2-y^2\right)\cdot\left[1+\sin\left(\omega t\right)\right],
\end{equation}
where $A$ is proportional to the pressure difference between the two ends of the channel, see fig.~\ref{fig:concept}. 
Here the heat effect caused by the dissipation process is ignored and assumed that the motion of the particle does not affect the velocity field of fluid for convenience. 

When the particle reaches the wall, we assume the velocity component perpendicular to the wall is zero while the velocity component parallel to the wall does not change. 
Regarding the orientation $\theta$, we assumed that it does not change upon collision (i.e., sliding boundary conditions~\cite{PhysRevLett.110.268301,PhysRevE.90.062301}). 
Note that this model can be regarded as a model of a nanorobot or other individuals moving in animal blood vessels since equation (\ref{eq:vrs_flow}) can be used as a simple approximation of the unidirectional and periodic flow of blood in blood vessels caused by heartbeat.

Next, we rewrite the Langevin dynamics by rescaling the length by the half wall width $r_0=Y$ and the time by relaxation time of Brownian motion $t_0=m/\gamma$. 
In this scaling the dimensionless equations of motion become
\begin{equation}\label{eq:nvrs_translation}
\frac{\mathrm d\boldsymbol v}{\mathrm dt}=a\boldsymbol n-\boldsymbol v+\boldsymbol B,
\end{equation}
\begin{equation}\label{eq:nvrs_rotation}
\frac{\mathrm d\boldsymbol n}{\mathrm dt}=c(\boldsymbol n\times\boldsymbol v)\times\boldsymbol n+\sqrt{2D}\xi(t)\;{\boldsymbol n}_{\boldsymbol\perp},
\end{equation}
where $\boldsymbol B=-b(1-\;y^2)\left[1+\sin(\Omega t)\right]{\boldsymbol e}_x$ and only five dimensionless parameters contained in the equations now:
\begin{equation}\label{eqs:nvrs_parameters}
a=\frac{mF}{Y\gamma^2},\;b=\frac{mAY}{2\gamma^2},\;c=\frac{Y\zeta}\tau,\;D=\frac{m\alpha}{\gamma\tau^2},\;\Omega=\frac{m\omega}\gamma.
\end{equation}

Note that parameter $a$ depends on the mass and thus can be regarded as a measure of inertia. The maximum speed of the background flow, the self-alignment effect, the strength of noise are characterized by parameters $b$, $c$ and $D$ respectively. 
To describe the anomalous diffusion regime we define the  effective anomalous diffusion coefficient
\begin{equation}\label{eq:def_D}
D_{\rm eff}=\frac{\left\langle\Delta x{(t)}^2\right\rangle}{2t^{\alpha}},
\end{equation}
and the power law grow for spread of particle trajectories in $x$-axis is obtained by
\begin{equation}\label{eq:def_alpha}
\alpha=t\frac{\mathrm d\ln\left\langle\mathrm\Delta x{(t)}^2\right\rangle}{\mathrm dt}, 
\end{equation}
where $\Delta x{(t)}=x{(t)}-x{(0)}$ and $\left\langle\cdot\right\rangle$ indicates averaging over all thermal noise realisations.

\section{\label{sec:sim}Result}
We simulate eq.~(\ref{eq:nvrs_translation}) and~(\ref{eq:nvrs_rotation}) using the Euler method with $c \equiv 1.0$, $\theta(0) = 0$ to find the diffusion regimes and the ultimate velocity in the $x$-direction $\left\langle v_x\left(t\right)\right\rangle$ and the probability distribution function in the $y$-direction.

\begin{figure}
\begin{center}
\includegraphics[width=\columnwidth]{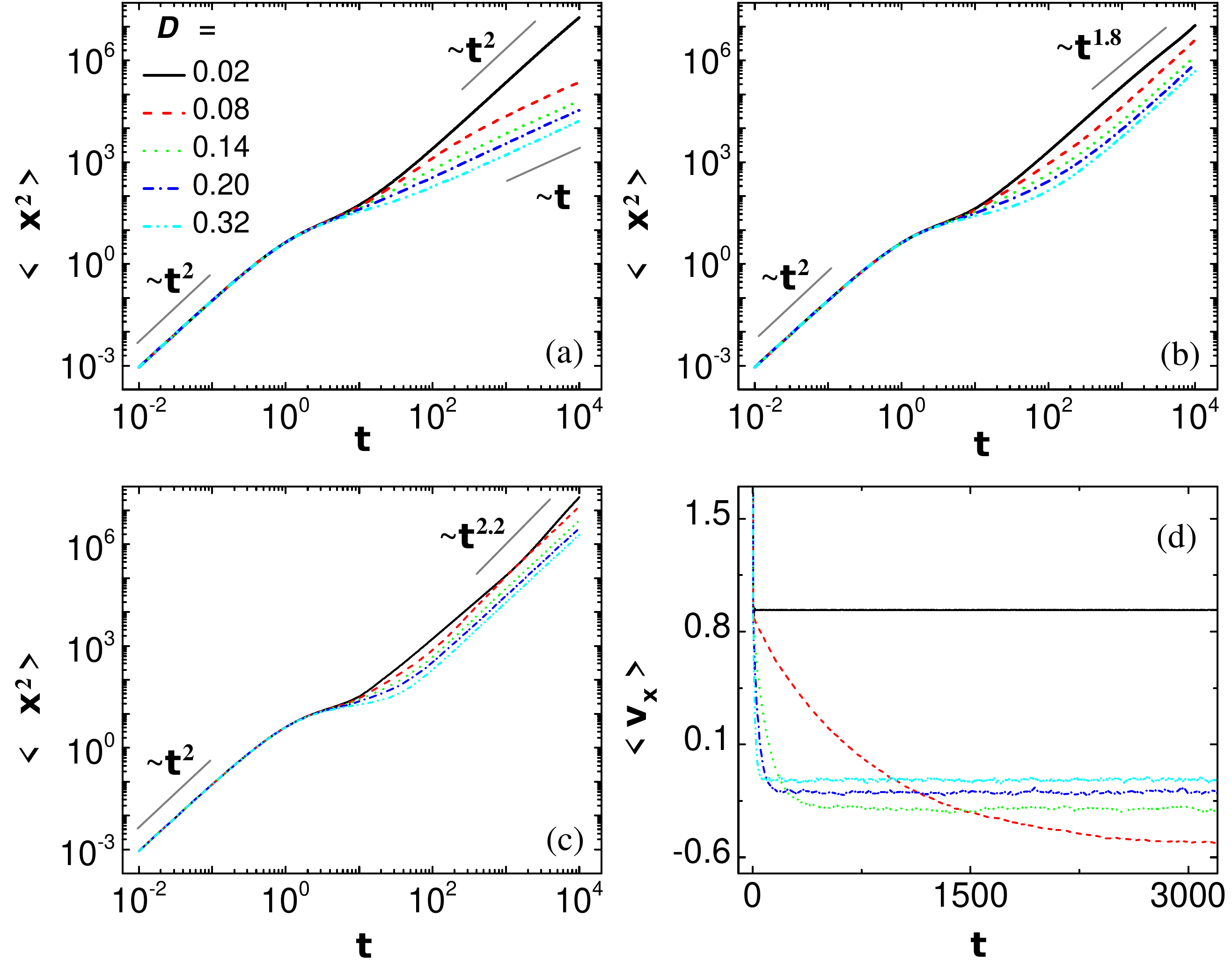}
\caption{Mean squared displacement and the velocity in the $x$-direction as a function of time. (a): $a=0.5$, $b=0$, shows the transition of normal diffusion to ballistic diffusion in the asymptotic long time limit. (b), $a=0.5$, $b=0.1$ shows superdiffusion regime and (c), $a=0.5$, $b=0.2$ superballistic diffusion in the asymptotic long time limit respectively. (d) Time evolution of averaged velocity for $a = 1$, $b = 0.1$. The frequency of the background flow $f=\Omega/2\pi$ equals $f=0$, colors represent the reduced rotational diffusion coefficient $D$, and gray solid lines are eye guides shows the linear behavior $\propto {t}^{\alpha}$.}
\label{fig:msd_v}
\end{center}
\end{figure}

In fig.~\ref{fig:msd_v} we depict the MSD($t$) as well as the velocity behavior as a function of time. 
We observe that particle moves ballistically ($\alpha = 2$) along the direction of initial velocity up to $t \approx 1$.  After the characteristic time, it starts to saturate and attain a diffusive scaling ($\alpha = 1$) up to $t \approx 10$. 
In fig.~\ref{fig:msd_v}(a), the MSD($t$) of the active particle moves in a static fluid background ($b=0$) are shown.
For $t \gg 1$ the MSD eventually recovers the diffusive scaling at high temperature (i.e., higher $D$ values) or return to ballistic diffusion regime at low temperature.  This different behavior is a direct consequence of the competition between rotation diffusion motion and self-aligning property. 

In fig.~\ref{fig:msd_v}(b), we find the superdiffusion scaling ($\alpha > 1$) in the limit of $t \gg 1$, which is typically related to the existence of broad probability densities of flights or strong correlations in the system~\cite{Babel_2014,PhysRevLett.93.120603,Hanes_2012,PhysRevE.88.032304}. 
Moreover, hyperdiffusion (superballistic diffusion, $\alpha > 2$) in the long time limit is observed in fig.~\ref{fig:msd_v}(c), which means that the coordinate variance grows faster than in the ballistic case, which was observed also for generalised Brownian motion in tilted washboard potential and was attributed to transient heating of particles from thermal bath, see~\cite{PhysRevE.76.061119,PhysRevLett.105.100602,PhysRevE.95.032107}. 
We note that the MSD of an ABPs is ballistic at short times and crosses over to a diffusive regime at long times, which is in stark contrast to the dynamics of self-align particles. 
The qualitative behavior of MSD of this two types of particle models is only consistent in the static flow background and high temperature, see fig.~\ref{fig:msd_v}(a).

To illustrate the competitive relationship between self-aligning property and rotational diffusion, the time evolution of averaged velocity for different $D$ values are plotted in fig.~\ref{fig:msd_v}(d). 
When the rotational diffusion coefficient $D$ is small (but does not have to be zero), average speaking, the change of the orientation of particle can almost be neglected and the particle moves up against the flow along the direction of the initial velocity. 
When $D$ becomes large, the orientation could change in a short time, and the velocity becomes negative due to the interaction with the background flow field in the negative direction, the reason why the speed of the station state gets slower with $D$ is that the diffusion process significantly affect the ballistic negative motion. 
Interestingly, the red line ($D=0.08$) plotted in fig.~\ref{fig:msd_v}(d) shows a long relaxation time to reach its stationary velocity, indicating a long-term competition between self-align and thermal noise couple with the flow background. 
  
\begin{figure}
\centering
\subfigure{
\includegraphics[width=\columnwidth]{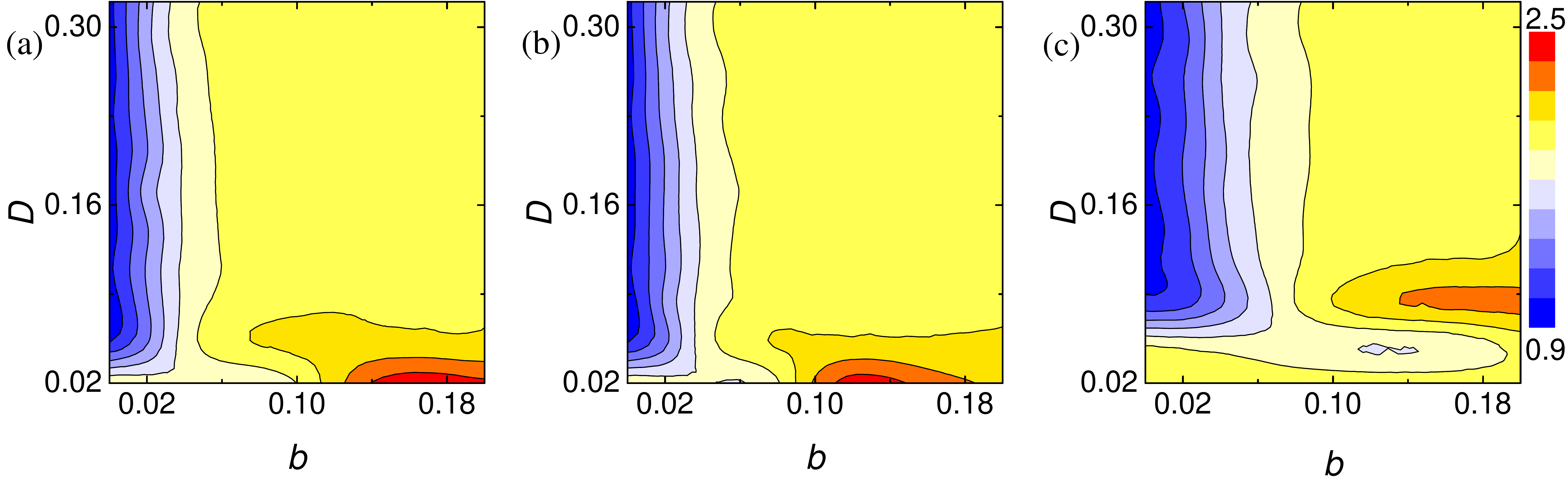}}
\subfigure{
\includegraphics[width=\columnwidth]{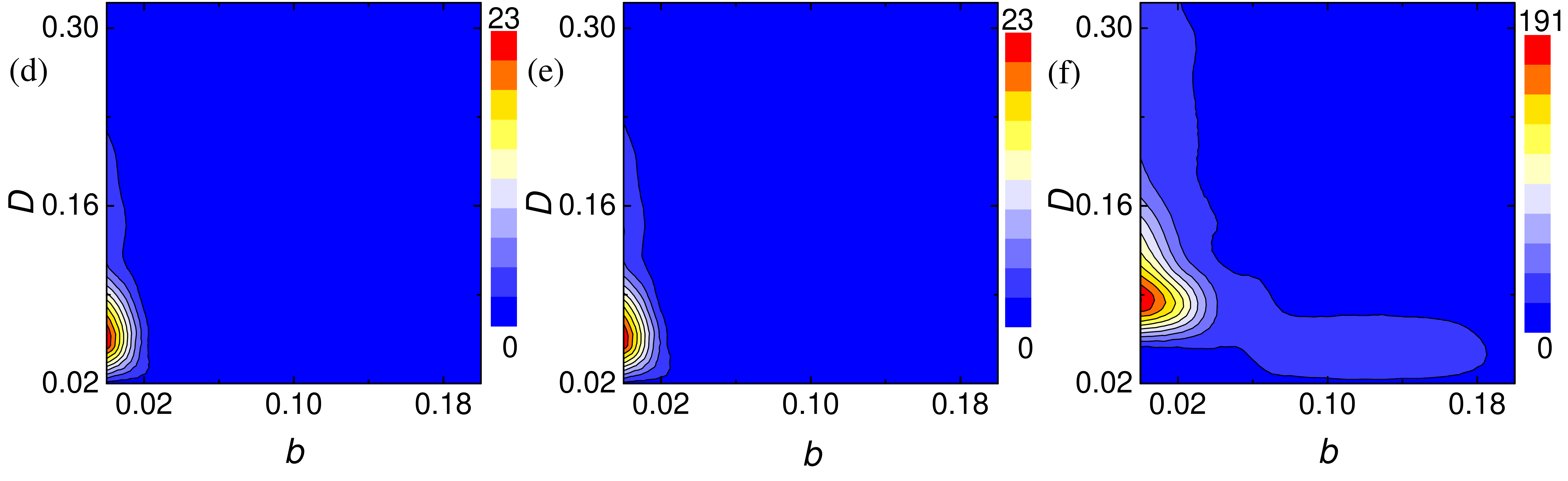}}
\caption{Heat maps showing the anomalous diffusion exponent $\alpha$ for panels (a)-(c) and effective anomalous diffusion coefficient $D_{\rm eff}$ for panels (d)-(f) calculated from simulation data as a function of $D$ and $b$. In panel (a) (d), $a=0.5$, $f=0$, in panel (b) (e) $a=0.5$, $f=0.01$ and $a=1.0$, $f=0$ for (c) (f).}
\label{fig:phase_alpha}
\end{figure}

By varying $D$ and $b$ (as well as frequency $f$), we find that the anomalous diffusion exponent $\alpha$ and the effective anomalous diffusion coefficient $D_{\rm eff}$ in the long time limit ($t \gg 1$) can be tuned over a wide range of values. 
For example, figs.~\ref{fig:phase_alpha}(a)-\ref{fig:phase_alpha}(c) shows the heat map of anomalous diffusion regime and figs.~\ref{fig:phase_alpha}(d)-\ref{fig:phase_alpha}(f) shows the corresponding effective anomalous diffusion coefficient for three parameters sets, respectively.
In the low flow velocity region (small $b$), if the rotational diffusion coefficient $D$ is small, the self-aligning property of the particle makes it maintain the original direction of motion, and then the ballistic diffusion/superdiffusion scaling is observed.
As temperature increases, the angular diffusion motion can overcome the effect of self-align and thus shows diffusion regime (see also fig.~\ref{fig:msd_v}(a) for example).
Since the velocity of the particle is determined by parameter $a$, and the self-align effect is increased with the velocity of the particle, the region of the ballistic diffusion/superdiffusion regime under low background flow velocity is increased with $a$ (compare with fig.~\ref{fig:phase_alpha}(a) and~ \ref{fig:phase_alpha}(c)).

When high background flow velocity (large $b$) is imposed, under the impact of background flow, the positive motion of particles slows down and the final motion direction of particles is always negative even for small (but not zero) $D$ values. 
If the rotational diffusion coefficient $D$ is small enough, the background fluid mainly accelerates the negative motion of particles, which explains the hyperdiffusion regions in figs.~\ref{fig:phase_alpha}(a)-\ref{fig:phase_alpha}(c) (see also figs.~\ref{fig:msd_v}(c)-\ref{fig:msd_v}(d)). 
As temperature increases, such negative motion will slow down (fig.~\ref{fig:msd_v}(d)) and the superdiffusion regime is recovered. 
The region of hyperdiffusion region is also different with $a$ since parameter $a$ can affect the self-propelled as well as the strength of the self-align effect, as discussed above.

Interestingly, in the diffusion regime (small $b$ value, $\alpha \approx 1$), we observed that the effective anomalous diffusion coefficient $D_{\rm eff}$ displays a resonance-like behavior (figs.~\ref{fig:phase_alpha}(d)-\ref{fig:phase_alpha}(f)). 
The peak value of $D_{\rm eff}$ and the location of the peak is sensitive to parameter $a$. 
Finally, we can see that the frequency of the background flow $f$ does not significantly affect the anomalous diffusion regime and $D_{\rm eff}$.

\begin{figure}
\begin{center}
\includegraphics[width=\columnwidth]{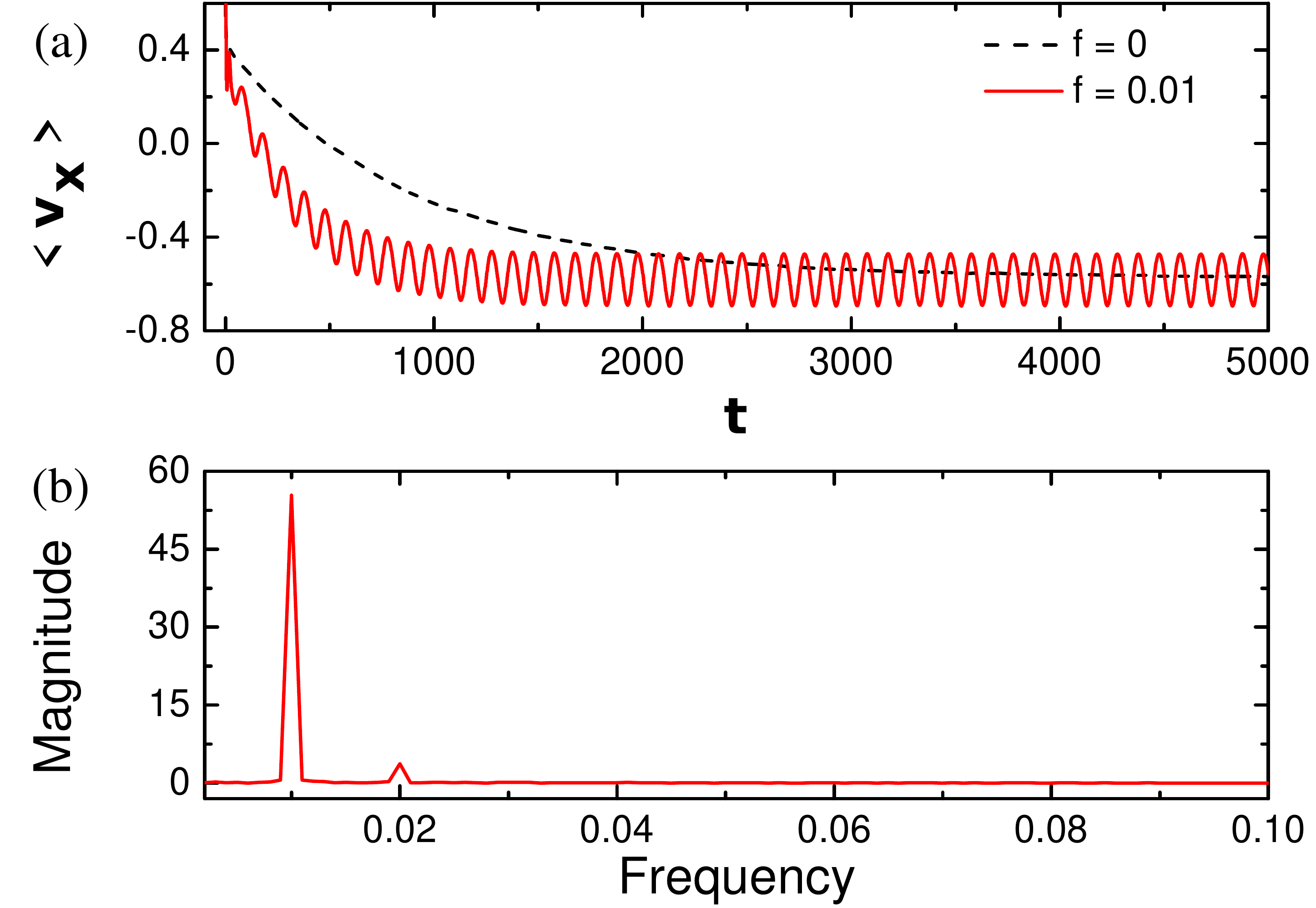}
\caption{(a) Evolution of averaged speed of particle with $a=0.5$, $b=0.2$, $D=0.02$. The result of Fourier analysis (b) proves that the oscillation of speed of the particle is a response to the background flow.}
\label{fig:relax_shorten}
\end{center}
\end{figure}

However, figure~\ref{fig:relax_shorten} shows another interesting phenomenon that if we impose a time-varying background flow, the relaxation time of particle reach the ``stationary'' velocity oscillation state can be shortened. 
The very long relaxation process for $f = 0$ is due to the long ``competition'' between self-align effect and rotational diffusion, which depend on the proper combination of parameters $a$, $b$, and $D$ (thus this phenomenon should only occur in a very small range of parameter sets). 
Therefore, when a periodic background fluid field is imposed (even the average speed is the same as that of steady flow), within a period of time when the flow speed deviates from the average value, the positive motion speed of active particle drops more rapidly. 
This can explain why a periodic background fluid could shorten the time that the particle reaches the stationary velocity oscillation state.

\begin{figure}
\begin{center}
\includegraphics[width=\columnwidth]{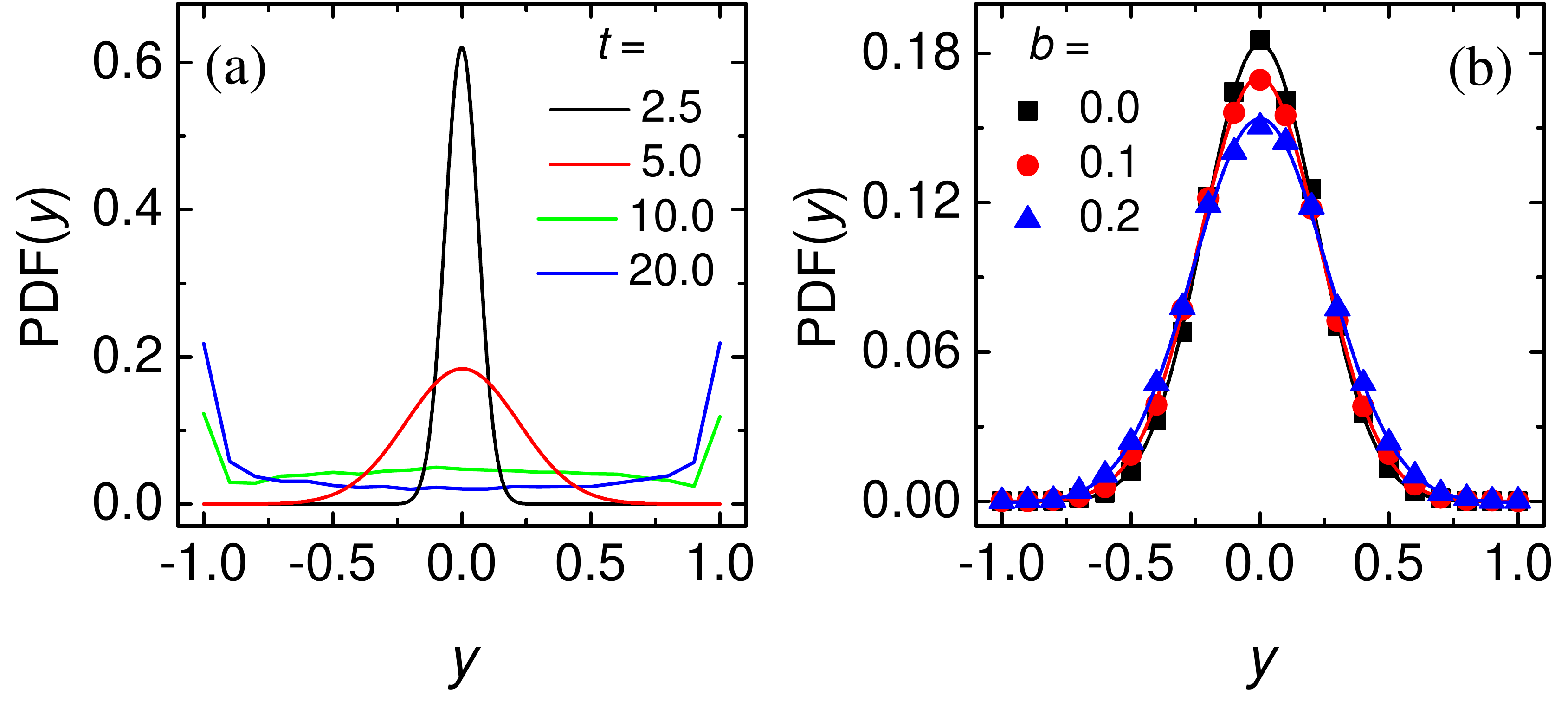}
\caption{Probability distribution functions of particles in the $y$ direction. (a) For $b=0.0$ at different time, (b) for $t=5.0$ at different $b$. Other parameters: $a=0.5$, $D=0.02$, $f=0$.}
\label{fig:y_distri}
\end{center}
\end{figure}

Figure \ref{fig:y_distri} describes the position probability distribution function (PDF) of particles in the $y$-direction in different scenarios. 
We observed that the PDF of particles in the $y$-direction presents a Gaussian distribution at the beginning, and forms a U-shaped distribution after reaching the boundary, i.e., active particles accumulate near boundaries and move along them, which has been widely observed~\cite{PhysRevLett.101.038102,PhysRevLett.123.178004,RevModPhys.88.045006}. 
For $b \neq 0$ (fig.~\ref{fig:y_distri}(b)) the Gaussian distribution of PDF(y) does not change, but the particle moves faster toward the boundary. 
This is because higher $b$ values weaken the self-align effect and the self-propelled direction of the particle is easier to deviate from the $x$-direction. 

\section{\label{sec:conclusion}Conclusion}
In summary, we explicitly considered an inertial active particle with self-aligning property moving in a Poiseuille flow. 
It is found that the effective anomalous diffusion coefficient changes sharply with temperature (i.e., parameter $D$) in the regime of low flow speed ($b \approx 0$). 
As the ratio of the mass of the particle, the self-propelled force and viscosity change, the type of long-time transportation regime could change significantly (from diffusive to hyperdiffusive scaling). 
Moreover, the relaxation time and ultimate average speed depending on the environmental diffusion coefficient non-monotonically, which are caused by the influence of $D$ on directional transport and the relationship between self-alignment and directional transport velocity. 
Last, the probability distribution functions of such a particle in the $y$-direction are Gaussian before interacting with the wall and become a U-shaped distribution after reaching the wall, we remark that this may be caused by our model design and has been observed in many experiments.

Since the rotational noise is positively correlated with the ambient temperature, and the behavior of the self-aligning particle under immobile fluid background with a not small $D$ is similar to the active particle without self-alignment (short-time interior ballistic motion, and then transformed into enhanced normal diffusion motion). 
Therefore, some active particles studied in previous works may possess self-aligning properties but have not been considered. 
Moreover, in the application of active matter, properties of self-propelled and self-alignment always are uncontrollable or unchangeable, while properties of the environment such as temperature, rotational diffusion coefficient, and even the direction (and the speed) of the background fluid are controllable (see~\cite{10.1038/s41467-020-17864-4} for example). 
Hence, it is fair to say that the results of this paper can provide a reference for the regulation of the movement of active substances in various environments.

\acknowledgments
We gratefully acknowledge support by the National Natural
Science Foundation of China (Grants No. 11975111 and
No. 12047501).

\bibliography{nvrs}

\end{document}